\documentclass[article, shortnames]{jss_arXiv}

\author{Sharon X. Lee\\University of Queensland \And 
        Geoffrey J. McLachlan\\University of Queensland}
\title{\pkg{EMMIX-uskew}: An \proglang{R} Package for Fitting 
Mixtures of Multivariate Skew $t$-distributions via the EM Algorithm}

\Plainauthor{Sharon  X. Lee, Geoffrey J. McLachlan} 
\Plaintitle{EMMIX-uskew: An R Package for Fitting Mixtures 
of Multivariate Skew $t$-distributions via the EM Algorithm} 
\Shorttitle{\pkg{EMMIX-uskew}: Fitting Mixtures of Multivariate Skew $t$-distributions} 

\Abstract{
  This paper describes an algorithm for fitting finite mixtures 
  of unrestricted Multivariate Skew $t$ (FM-uMST) distributions. 
  The package \pkg{EMMIX-uskew} implements a closed-form 
  expectation-maximization (EM) algorithm 
  for computing the maximum likelihood (ML) estimates of the 
  parameters for the (unrestricted) FM-MST model in \proglang{R}. 
  \pkg{EMMIX-uskew} also supports visualization of fitted contours 
  in two and three dimensions, and random sample generation 
  from a specified FM-uMST distribution.  
	 
  Finite mixtures of skew $t$-distributions have proven to be useful in modelling 
  heterogeneous data with asymmetric and heavy tail behaviour, 
  for example, datasets from flow cytometry. In recent years, 
  various versions of mixtures with
  multivariate skew $t$ (MST) distributions
  have been proposed. However, these models adopted some 
  restricted characterizations of the component MST distributions so that
  the E-step of the EM algorithm can be evaluated in closed form.
  This paper focuses on mixtures with unrestricted MST components,
  and describes an iterative algorithm for the computation
  of the ML estimates of its model parameters.
  Its implementation in \proglang{R} is presented 
  with the package \pkg{EMMIX-uskew}.  
	
	The usefulness of the proposed algorithm is demonstrated in three  		
	applications to real datasets.  The first example illustrates the 
	use of the main function \proglang{fmmst} in the package by fitting
	a MST distribution to a bivariate unimodal flow cytometric sample. 
	The second example fits a mixture of MST distributions to 
	the Australian Institute of Sport (AIS) data, and demonstrates that 
	\pkg{EMMIX-uskew} can provide better clustering results than 
	mixtures with restricted MST components.  	
	In the third example, \pkg{EMMIX-uskew} is applied to classify cells 
	in a trivariate flow cytometric dataset.
  Comparisons with some other available methods suggest that 
  \pkg{EMMIX-uskew} achieves a lower misclassification rate 
  with respect to the labels given by benchmark gating analysis.
}
\Plainkeywords{mixture models, skew distributions, multivariate t-distribution, EM algorithm, flow cytometry, R} 

\Address{
  Geoffrey J. McLachlan\\
  Department of mathematics\\
  University of Queensland\\
  Brisbane, Australia\\
  E-mail: \email{g.mclachlan@uq.edu.au}\\
}

\usepackage{graphicx,amssymb,amstext,amsmath, amsfonts, breakurl}

\newcommand{\btheta}  {\boldsymbol\theta}
\newcommand{\bmu}     {\boldsymbol\mu}

\newcommand{\bgamma}  {\boldsymbol\gamma}
\newcommand{\bdelta}  {\boldsymbol\delta}
\newcommand{\bPsi}    {\boldsymbol\Psi}
\newcommand{\bLambda} {\boldsymbol\Lambda}

\newcommand{\bSigma}  {\boldsymbol\Sigma}
\newcommand{\bDelta}  {\boldsymbol\Delta}
\newcommand{\bOmega}  {\boldsymbol\Omega}
\newcommand{\be}	  {\mbox{\boldmath $e$}}
\newcommand{\bq}	  {\mbox{\boldmath $q$}}
\newcommand{\bs}	  {\mbox{\boldmath $s$}}
\newcommand{\bu}	  {\mbox{\boldmath $u$}}
\newcommand{\by}	  {\mbox{\boldmath $y$}}
\newcommand{\bz}	  {\mbox{\boldmath $z$}}
\newcommand{\bI}	  {\mbox{\boldmath $I$}}
\newcommand{\bU}	  {\mbox{\boldmath $U$}}
\newcommand{\bS}	  {\mbox{\boldmath $S$}}
\newcommand{\bX}	  {\mbox{\boldmath $X$}}
\newcommand{\bY}	  {\mbox{\boldmath $Y$}}
\newcommand{\bZ}	  {\mbox{\boldmath $Z$}}
\newcommand{\byj}     {\by_j}
\newcommand{\kth}     {^{(k)}}
\newcommand{\qhjk}    {\boldsymbol q_{hj}\kth}
\newcommand{\dhkyj}   {d_h\kth(\byj)}
\newcommand{\bzero}	  {\mbox{\boldmath $0$}}

\begin{document}


\section{Introduction}

In many practical problems, data are often skewed, heterogeneous, 
and/or contain outliers. 
Finite mixture distributions 
of skewed distributions have become increasingly popular 
in modelling and analyzing such data. 
This use of finite mixture distributions to model 
heterogeneous data has undergone intensive development in 
the past decades, as witnessed by the numerous applications 
in various scientific fields such as bioinformatics, 
cluster analysis, genetics, information processing, 
medicine, and pattern recognition. 
For a comprehensive survey on mixture models and 
their applications see, for example, 
the monographs by \citet{B013}, \citet{B008}, 
\citet{B006}, \citet{B011}, \citet{B014}, \citet{B003},
and \citet{B015}, the edited volume of \citet{B016},
and also the papers by \citet{J059} and \citet{J060}.

In recent years, finite mixtures of skew $t$-distributions
have been exploited as an effective tool in modelling 
high-dimensional multimodal and asymmetric datasets; 
see, for example, \citet{J004} and \citet{J028}.
Following the introduction of the skew normal (SN) distribution
by \citet{J005}, several authors have studied skewed extensions 
of the $t$-distribution. 
Finite mixture models with multivariate skew $t$ (MST) components 
was first proposed by \citet{J004} in a study of 
an automated approach to the analysis of flow cytometry data.
\citet{S002} has given a package \pkg{EMMIX-skew} 
for the implementation in \proglang{R} \citep{S004} of their algorithm.
More recently, \citet{J047} studied a class of mixture models where 
the components densities are scale mixtures of univariate skew normal 
distributions, 
known as the skew normal/independent (SNI) family of distributions, 
which include the (univariate) skew normal and skew 
$t$-distributions as special cases. 
This work was later extended to the multivariate case in
\citet{J066}, and was implemented in an \proglang{R} package
\pkg{mixsmsn}. However, in these characterizations, 
restrictions were imposed on the component skew $t$-distributions
in order to obtain manageable analytical expressions 
for the conditional expectations involved in the E-step
of the EM algorithm. 
These versions of the skew $t$-distributions are known as 
the `restricted' form of the MST distribution; 
see \citet{J103} for further discussion on this.

In this paper, we present an algorithm for the fitting of
the unrestricted skew $t$-mixture model. We show that
an EM algorithm can be implemented exactly 
without restricting the characterizations 
of the component MST distributions. 
Closed form expressions can be obtained for the 
E-step conditional expectations by recognizing that
they can be formulated as moments of a multivariate 
non-central truncated $t$-variate, which can be further 
expressed in terms of central $t$-distributions.
The algorithm is implemented in \proglang{R} 
in the package \pkg{EMMIX-uskew}, available at
\url{http://www.maths.uq.edu.au/~gjm/mix_soft/EMMIX-skew}. 

The package \pkg{EMMIX-uskew} consists of three main functions:
\proglang{fmmst}, \proglang{rfmmst}, and \proglang{contour.fmmst}. 
The main function \proglang{fmmst} fits a mixture of unrestricted MST (uMST)
distributions using an EM algorithm described in 
Section~\ref{sec2}. The function \proglang{rfmmst} 
generates random samples from mixtures of uMST distributions.
For a user friendly visualisation of the fitted models, 
\proglang{fmmst.contour} provides 2D contour maps of 
the fitted bivariate densities and 3D displays with 
interactive viewpoint navigation facility for trivariate densities.

The remainder of this paper is organized as follows. 
Section~\ref{sec1} provides a brief description of the 
uMST distribution and defines the FM-uMST model.
Section~\ref{sec2} presents an EM algorithm for fitting
the FM-uMST model. 
In the next section, an explanation of how to fit, 
visualize, and interpret the FM-uMST models using \pkg{EMMIX-uskew} is presented. 
The usage of \pkg{EMMIX-uskew} is illustrated with three applications 
and comparisons are made with some restricted FM-MST models 
and other clustering methods. 
Finally, we conclude with a brief summary of our results.

\section{Finite mixtures of multivariate skew $t$-distributions}
\label{sec1}

We begin by defining the (unrestricted) multivariate skew $t$-density.
Let $\bY$ be a $p$-dimensional random vector. 
Then $\bY$ is said to follow a $p$-dimensional 
unrestricted skew $t$-distribution \citep{J002} with 
$p \times 1$ location vector $\bmu$, 
$p \times p$ scale matrix $\bSigma$, 
$p \times 1$ skewness vector $\bdelta$, 
and (scalar) degrees of freedom $\nu$, 
if its probability density function (pdf) 
is given by 
\begin{equation}
f_p(\boldsymbol y; \bmu, \bSigma, \bdelta, \nu) 
		= 2^p t_{p, \nu}\left(\by; \bmu, \bOmega\right) 
		T_{p, \nu+p}\left(\by^*; \boldsymbol 0, \bLambda\right),
\label{eq:MST}
\end{equation}
where 
\begin{align} 
\bDelta &= \mbox{diag}(\bdelta), \notag\\
\bOmega &= \bSigma + \bDelta^2, \notag\\
\by^* &= \bq \sqrt{\frac{\nu+p}{\nu+d\left(\by\right)}}, \notag\\
\bq &= \bDelta\bOmega^{-1} (\by-\bmu), \notag\\
d\left(\by\right) &= (\by-\bmu)^T \bOmega^{-1}(\by-\bmu), \notag\\
\bLambda &= \bI_p - \bDelta \bOmega^{-1} \bDelta. \notag
\end{align}
Here the operator $\mbox{diag}(\bdelta)$ denotes 
a diagonal matrix with diagonal elements specified by 
the vector $\bdelta$. 
Also, we let $t_{p, \nu}(.; \bmu, \bSigma)$ be the 
$p$-dimensional $t$-distribution with 
location vector $\bmu$, scale matrix $\bSigma$, 
and degrees of freedom $\nu$, 
and $T_{p,\nu} (.;\bmu, \bSigma)$ the corresponding 
(cumulative) distribution function. 
The notation $\bY \sim \mbox{uMST}_{p, \nu}
(\bmu, \bSigma, \bdelta)$ will be used. 
Note that when $\bdelta = \boldsymbol 0$, (\ref{eq:MST}) 
reduces to the symmetric $t$-density 
$t_{p,\nu}(\by; \bmu, \bSigma)$. 
Also, when $\nu \rightarrow \infty$, 
we obtain the (unrestricted) skew normal distribution. 

Various versions of the multivariate skew $t$-density
have been proposed in recent years. 
It is worth noting that the versions considered by 
\citet{J006}, \citet{J019}, and \citet{J049}, 
among others, are different from (\ref{eq:MST}). 
These versions are simpler in that the 
skew $t$-density is defined in terms 
involving only the univariate $t$-distribution function 
instead of the multivariate form of the latter 
as used in (\ref{eq:MST}). 
These simplified characterizations have the 
advantage of having closed form expressions
for the conditional expectations 
that have to be calculated on the E-step.  
The reader is referred to \citet{J103, J105} 
for a discussion on different forms of skew $t$-distributions. 
We shall adopt the unrestricted form (\ref{eq:MST}) of the MST distribution here
as proposed by \citet{J002}, and describe 
a computationally efficient EM algorithm for fitting this model.

A $g$-component finite mixture of uMST distributions
has density given by
\begin{equation}
f\left(\by; \bPsi\right) = \sum_{h=1}^g \pi_h f_p 
	\left(\by; \bmu_h, \bSigma_h, \bdelta_h , \nu_h \right),
\label{eq:MSTmix} 
\end{equation}
where $f_p \left(\by; \bmu_h, \bSigma_h, \bdelta_h, \nu_h \right)$ 
denotes the $h$th uMST component of the mixture model 
as defined by (\ref{eq:MST}), with location parameter $\bmu_h$, 
scale matrix $\bSigma_h$, skew parameter $\bdelta_h$, 
and degrees of freedom $\nu_h$. 
The mixing proportions $\pi_h$ satisfy 
$\pi_h \geq 0$ $(h=1,\ldots,g)$ and 
$\sum_{h=1}^g \pi_h = 1$. 
We shall denote the model defined by (\ref{eq:MSTmix}) by 
the FM-uMST (finite mixture of uMST) distributions. 
Let $\bPsi$ contain all the unknown parameters of 
the FM-uMST model; that is, 
$\bPsi = \left(\pi_1, \ldots, \pi_{g-1}, \btheta_1^T, \ldots, 
\btheta_g^T\right)^T$ where now $\btheta_h$ consists of 
the unknown parameters of the $h$th component density function.
The density values for a uMST and FM-uMST distribution can be 
evaluated using the functions \code{dmst} and \code{dfmmst}
in \pkg{EMMIX-uskew}.

Random samples of uMST variates can be generated by adopting 
a stochastic representation of (\ref{eq:MST}) \citep{J027}. 
If $\bY \sim uMST_{p,\nu}(\bmu, \bSigma, \bdelta)$, then 
\begin{equation}
\bY = \bmu + \frac{1}{\sqrt{w}}\bDelta \left|\bU_1\right| 
		+ \frac{1}{\sqrt{w}} \bU_0,
\label{eq2.1}
\end{equation}
where the random variables
\begin{align}
\bU_0 &\sim 	N_p(\boldsymbol 0, \bSigma), \label{eq2.2}\\
\bU_1 &\sim	N_p(\boldsymbol 0, \bI_p), \label{eq2.3}\\
w	&\sim	\mbox{gamma} \left(\frac{\nu}{2}, \frac{\nu}{2}\right), \label{eq2.4}
\end{align}
are independent, and $\mbox{gamma}(\alpha,\beta)$ 
denotes the gamma distribution 
with shape and scale parameters 
given by $\alpha$ and $\beta$ respectively. 
Sampling of uMST and FM-uMST variates are implemented in
\pkg{EMMIX-uskew} in the \code{rmst} and \code{rfmmst} functions,
respectively.

\section{The EMMIX-uskew algorithm}
\label{sec2}

From (\ref{eq2.1}) to (\ref{eq2.4}), 
the uMST distribution admits a convenient hierarchical 
characterization that facilitates the computation of the 
maximum likelihood estimator (MLE) of the unknown 
model parameters using the EM algorithm, namely,
\begin{eqnarray}
\bY \mid \bu, w &\sim&	
		N_p\left(\bmu + \bDelta \bu, \frac{1}{w} \bSigma\right), 
		\nonumber\\
\bU \mid w &\sim&	
		HN_p\left(\boldsymbol 0, \frac{1}{w} \bI_p\right), \nonumber\\
W	&\sim&	\mbox{gamma}\left(\frac{\nu}{2}, 
		\frac{\nu}{2}\right), \nonumber
\label{eq:MSTMIX_H}
\end{eqnarray} 
where $\bDelta_h = \mbox{diag}\left(\bdelta_h\right)$, 
$HN_p(\bmu, \bSigma)$ denotes the $p$-dimensional half-normal 
distribution with location parameter $\bmu$ and scale matrix 
$\bSigma$. 

\subsection{Fitting of FM-uMST model via the EM algorithm}
\label{sec2.1}
 
Let $\bY_1, \cdots, \bY_n$ be $n$ independent observations of $\bY$.
To formulate the estimation of the unknown parameters as 
an incomplete-data problem in the EM framework, 
we introduce a set of latent component labels 
$\bz_j = (z_{1j}, \ldots, z_{gj})$ $(j=1,\ldots,n)$ 
in addition to the unobservable variables $\bu_j$ and $w_j$, 
where each element $z_{hj}$ is a zero-one indicator variable 
with $z_{hj}=1$ if $\by_j$
belongs to the $h$th component, and zero otherwise. 
Thus, $\sum_{h=1}^g z_{hj} =1$ $(j=1,\ldots,n)$. 
It follows that the random vector $\boldsymbol Z_j$ 
corresponding to $\boldsymbol z_j$ follows a multinomial 
distribution with one trial and cell probabilities 
$\pi_1, \ldots, \pi_g$; that is, 
$\bZ_j \sim \mbox{Mult}_g (1; \pi_1, \ldots,\pi_g)$.

The complete-data log likelihood function can be factored into
the marginal densities of the $\bz_j$, the conditional densities of 
the $w_j$ given $\bz_j$, and the conditional densities of 
the $\by_j$ given $\bu_j$, $w_j$, and $\bz_j$. 
Accordingly, the complete-data log likelihood is given by
\begin{equation}
\log L_c\left(\bPsi\right) = \log L_{1c}\left(\bPsi\right) + 
\log L_{2c}\left(\bPsi\right) + \log L_{3c}\left(\bPsi\right),
\label{logL}
\end{equation}
where
\begin{align}
L_{1c}\left(\bPsi\right) &=	\sum_{h=1}^g \sum_{j=1}^n z_{hj} 
\log\left(\pi_h\right), \displaybreak[0]\notag\\
L_{2c}\left(\bPsi\right) &=	\sum_{h=1}^g \sum_{j=1}^n z_{hj} 
\left[\left(\frac{\nu_h}{2}\right) \log\left(\frac{\nu_h}{2}\right) + 
\left(\frac{\nu_h}{2}+p-1\right)\log\left(w_j\right) \right.\notag\\
	&		\left .- \log\Gamma\left(\frac{\nu_h}{2}\right) - 
	\left(\frac{w_j}{2}\right)\nu_h\right], \notag\displaybreak[0]\\
L_{3c}\left(\bPsi\right) &=	\sum_{h=1}^g \sum_{j=1}^n z_{hj} \left\{- 
p\log\left(2\pi\right) -  \frac{1}{2}\log\left|{\bSigma}_h\right| \right. 
\notag\\
	&		- \left. \frac{w_j}{2}\left[d_h\left(\byj\right) + 
	\left(\bu_j-\boldsymbol q_{hj}\right)^T \bLambda_h^{-1} 
	\left(\bu_j-\boldsymbol q_{hj}\right)\right]\right\}, 
\label{Lc2}
\end{align}
and where 
\begin{align}
d_h\left(\byj\right) &= \left(\byj-\bmu_h \right)^T {\bOmega}_h^{-1} 
\left(\byj-\bmu_h\right), \notag\\
\boldsymbol q_{hj} &= \bDelta_h {\bOmega}_h^{-1} 
\left(\byj-\bmu_h\right), \notag\\ 
{\bLambda}_h &= \boldsymbol I_p - \bDelta_h{\bOmega}_h^{-1}\bDelta_h, 
\notag\\
{\bOmega}_h &= {\bSigma}_h + \bDelta_h^2. \notag
\end{align}
Here $\bPsi$ contains all the unknown parameters of the FM-uMST model.
    
The implementation of the EM algorithm requires alternating 
repeatedly the E- and M-steps until convergence 
in the case where the changes in the log likelihood values 
are less than some specified small value. 
The E-step calculates the expectation of the complete-data 
log likelihood given the observed data $\by$ using the 
current estimate of the parameters, known as the $Q$-function, 
given by
\begin{eqnarray}
Q(\bPsi; \bPsi^{(k)}) &=&	E_{\bPsi^{(k)}}
		\left\{\log L_c\left(\bPsi\right)
		 \mid \by\right\}. \nonumber
\end{eqnarray}

The M-step then maximizes the $Q$-function 
with respect to the parameters $\bPsi$. 

On the $(k+1)$th iteration, the E-step requires the calculation
of the conditional expectations
\begin{align}
e_{1,j}^{(k)} &= E_{\btheta^{(k)}} \left(W_j\mid\byj \right), 
\label{eq1}\\ 
\be_{2,j}\kth &= E_{\btheta^{(k)}} \left(W_j\bU_j\mid\byj\right), 
\label{eq2}\\ 
\be_{3,j}^{(k)}&=E_{\btheta^{(k)}} \left(W_j\bU_j\bU_j^T \mid \byj\right). 
\label{eq3}
\end{align}

The conditional expectation of $Z_{hj}$ given the observed data, is given, 
using Bayes' Theorem, by
\begin{equation}
\tau_{hj}^{(k)} = \frac{\pi_h^{(k)} f_p \left(\byj; 
		\bmu_h^{(k)}, {\bSigma}_h^{(k)}, \bdelta_h^{(k)}, 
		\nu_h\kth\right)} 
		{\sum_{i=1}^g \pi_i^{(k)} f_p \left(\byj;\bmu_i^{(k)}, 
		{\bSigma}_i^{(k)}, \bdelta_i^{(k)}, \nu_i\kth\right)}.
\label{eq:TAU}
\end{equation} 
which can be interpreted as the posterior probability of membership of the $h$th 
component by $\byj$, using the current estimate $\bPsi^{(k)}$ for $\bPsi$.

It can be shown that the conditional expectations 
$e_{1,j}^{(k)}$, $\be_{2,j}^{(k)}$, and $\be_{3,j}^{(k)}$
are given by
\begin{align}
e_{1,hj}^{(k)} 
	&=\left(\frac{\nu_h^{(k)}+p}
	{\nu_h^{(k)}+d_h^{(k)}\left(\byj\right)}\right) 		
	\frac{T_{p,\nu_h^{(k)}+p+2}\left(\qhjk 
	\sqrt{\frac{\nu_h^{(k)}+p+2}{\nu_h^{(k)}+\dhkyj}};
	\boldsymbol 0, 	{\bLambda}_h^{(k)}\right)} 
	{T_{p,\nu_h^{(k)}+p}\left(y_{hj}^{*(k)}; \boldsymbol 0,
	{\bLambda}_h^{(k)}\right)}, 
	\label{eq:e2} \displaybreak[0]\\
\be_{2,hj}^{(k)} 	&= 
		e_{1,hj}^{(k)} E(\bX), 		
		\label{eq:e3} \displaybreak[0]\\
\intertext{and} 
\be_{3,hj}^{(k)} 	&=
		e_{1,hj}^{(k)} E(\bX \bX^T), 
		\label{eq:e4} \displaybreak[0]
\end{align} 
where $\boldsymbol X$ is a $p$-dimensional $t$-variate truncated 
to the positive hyperplane $\mathbb{R}^+$, 
which is distributed as
\begin{equation}
\bX \sim tt_{p,\nu_h^{(k)}+p+2}
		\left(\boldsymbol q_{hj}^{(k)},
		\left(\frac{\nu_h^{(k)}+d_h^{(k)}(\byj)} {\nu_h^{(k)}+p+2}\right) 
		\bLambda_h^{(k)}; \mathbb{R}^+\right),
\label{eqX}
\end{equation} 
where $tt_{p, \nu}(\bmu, \bSigma; \mathbb{R}^+)$ denotes the 
positively truncated $t$-distribution with location vector $\bmu$, 
scale matrix $\bSigma$, and $\nu$ degrees of freedom.
The truncated moments $E(\bX)$ and $E(\bX \bX^T)$ 
can be swiftly evaluated by noting that 
they can be expressed in terms of the distribution function of 
a (non-truncated) multivariate central $t$-random vector; \citet{J068, J103}. 
Recently, \citet{J067} have considered the moments of of the doubly truncated 
multivariate $t$-distribution, but their result corresponding to (\ref{eqX}) 
is incorrect; see \citet{J103} for further details.

The $(k+1)$th M-step consists of maximization of the $Q$-function
with respect to $\bPsi$. It follows that an updated estimate of the
unknown parameters of the FM-uMST model is given by 
\begin{align}
\bmu_h^{(k)} &= \frac{\sum_{j=1}^n 
	\tau_{hj}^{(k)} \left[e_{1,hj}^{(k)}\byj - 
	\bDelta_h^{(k)}\be_{2,hj}^{(k)}\right]} {\sum_{j=1}^n 
	\tau_{hj}^{(k)}e_{1,hj}^{(k)}}, 
	\label{eq:mu}\displaybreak[0]\\
		\bdelta_h^{(k+1)} &= 
		\left(\bSigma_h^{(k)^{-1}} \odot \sum_{j=1}^n \tau_{hj}^{(k)} 
		\be_{3,hj}^{(k)}\right)^{-1} \mbox{diag} 
		\left(\bSigma_h^{(k)^{-1}} \sum_{j=1}^n \tau_{hj}^{(k)} 
		\left(\by_j - \bmu_h^{(k+1)}\right) \be_{2,hj}^{(k)^T}\right), 
	\label{eq:delta}\displaybreak[0]\\%
	\intertext{and}
		{\bSigma}_h^{(k+1)} &=	\frac{1} {\sum_{j=1}^n 
		\tau_{hj}^{(k)}} \sum_{j=1}^n \tau_{hj}^{(k)} 
		\left[\bDelta_h^{(k+1)}\be_{3,hj}^{(k)^T}\bDelta_h^{(k+1)^T} 
		 - \left(\byj-\bmu_h^{(k+1)}\right) \be_{2,hj}^{(k)^T} 
	\bDelta_h^{(k+1)} \right. \notag\\
	&	- \left.  \bDelta_h^{(k+1)}\be_{2,hj}^{(k)} 
	\left(\byj-\bmu_h^{(k+1)}\right)^T 
		+ \left(\byj-\bmu_h^{(k+1)}\right) 
	\left(\byj-\bmu_h^{(k+1)}\right)^T 
	e_{1,hj}^{(k)}\right], 
	\label{eq:sigma}
\end{align}
where $\odot$ denotes element-wise matrix product.
Note that (\ref{eq:delta}) and also (\ref{eqX}) 
are given incorrectly in \citet{J068}.

An update $\nu_h^{(k+1)}$ of the degrees of freedom 
is obtained by solving iteratively the equation
\begin{equation}
		\log\left(\frac{\nu_h^{(k+1)}}{2}\right) 
	- \psi\left(\frac{\nu_h^{(k+1)}}{2}\right) 
	= \frac{\sum_{j=1}^n 	\tau_{hj}^{(k)} \left[
	\log\left(\frac{\nu_h^{(k)} + d_h^{(k)}(\by_j)}{2}\right)
	- \psi\left(\frac{\nu_h^{(k)}+p}{2}\right)
	+ \frac{\nu_h^{(k)}+p}{\nu_h^{(k)}+d_h^{(k)}(\by_j)}\right]} 
	{\sum_{j=1}^n \tau_{hj}^{(k)}},
	\nonumber
\end{equation} 
where $\psi(x) = \frac{\Gamma'(x)}{\Gamma(x)}$ is the Digamma function. 
This last equation has been simplified by making use of a one-step-late 
approximation \citep{J061} in updating the estimate of $\nu_h$.
As a consequence, it can affect the monotonicity of the likelihood function.
Our experience suggests that this rarely happens.
The monotonicity of the likelihood can be preserved 
by working with the exact expression
as given by Equation (73) in \citet{J103}. 
There is an option in the program to use this more time consuming 
updating of $\nu_h$. 
The algorithm described in this Section is implemented 
as the \code{fmmst} function in \pkg{EMMIX-uskew}. 
\newline

\subsection{Choosing initial values}
\label{sec2.2}

It is important to obtain suitable initial values 
in order for \code{fmmst} to converge quickly.
In \pkg{EMMIX-uskew}, starting values for the model parameters 
are based on an initial clustering given by $k$-means.
Twenty attempts of $k$-means are performed, and 
the starting component labels $\bz_j^{(0)}$ $(j=1,\ldots,n)$
are initialized according to the clustering result 
with the highest relative log likelihood (see \citet{J103}). 
The other parameters are initialized as follows:
\begin{eqnarray}
\bSigma^{(0)}	&=& \bS_h - (a-1) \; 	
	\mbox{diag}\left(\bs_h\right),
	\nonumber\\
\bdelta^{(0)} &=& \mbox{sign}(\bgamma_h) \sqrt{\frac{(1-a)\pi}{\pi-2}}
	\; \bs_h^*,
	\nonumber\\
\bmu^{(0)}	&=&	\bar{\by} - \sqrt{\frac{2}{\pi}} \bdelta^{(0)},
	\nonumber\\	
\nu^{(0)} &=&	40, 	
\label{eq:init}
\end{eqnarray}
where $\bS_h$ is the sample covariance of the $h$th component,
and where $\bgamma_h$ is the sample skewness of the $h$th component, 
whose $i$th element is given by
\begin{eqnarray}
\gamma_i &=&	\frac{n^{-1} \sum_{j=1}^n (y_{ij}-\mu_i)^3} 
		{\left(n^{-1} \sum_{j=1}^n (y_{ij}-\mu_i)^2\right)
		^{\textstyle\frac{3}{2}}} 
		\;\;\;\; (i=1,\,\ldots,\,p),
		\nonumber
\end{eqnarray}
and where $y_{ij}$ denotes the $i$th element of the $j$th observation,
and $\mu_i$ is the $i$th element of $\bmu$.  
Here, $\bs_h$ denotes the vector created by
extracting the main diagonal of $\bS_h$, 
and the vector $\bs_h^*$ is created by 
taking the square root of each element in $\bs_h$. 
The scalar $a$ is varied systematically across the interval $(0,1)$ 
to search for a (relatively) optimal set of starting values for the model parameters.

\subsection{Stopping rule}
\label{sec2.3}

\pkg{EMMIX-uskew} adopts a traditional stopping criterion which is 
based on the absolute change in the size of the log likelihood. 
An Aitken acceleration-based strategy is described in \cite{J027}.
The algorithm is terminated when the absolute difference 
between the log likelihood value and the asymptotic log likelihood value
is less than a sepcified tolerance, $\epsilon$, that is   
\begin{equation}
\left|L_\infty^{(k+1)} - L^{(k+1)}\right| < \epsilon,
\label{eq:stop2}
\end{equation}
where $L_\infty^{(k+1)}$ is the asymptotic estimate of the log likelihood 
at the $(k+1)$th iteration, given by $L_\infty^{(k+1)} = L^{(k)} + \frac{L^{(k+1)} - L^{(k)}} {1-\alpha^{(k)}}$, 
and $\alpha^{(k)} = \frac{L^{(k+1)}-L^{(k)}} {L^{(k)}-L^{(k-1)}}$ is the Aitken's acceleration at the $k$th iteration.
The default tolerance is $\epsilon = 10^{-3}$, 
but the user can specify a different value.

\section{Using the EMMIX-uskew package}
\label{sec3}

The parameters of the FM-uMST model in \pkg{EMMIX-uskew} are specified as
a list structure containing the elements 
described in Table~\ref{tab1}. The parameters $\bmu$, $\bSigma$, 
and $\bdelta$ are each implemented as a list of \code{g} matrices, 
where \code{g} is the number of components in the fitted model. 
For example, \code{mu[[2]]} is a $p \times 1$ matrix 
representing $\bmu_2$.  
Each \code{sigma[[h]]} $(h=1,\ldots,g)$ is a $p\times p$ matrix 
representing the symmetric positive definite scale matrix of 
the $h$th component. The parameters \code{dof} and \code{pro}
are $g$ by $1$ arrays, representing the vector of degrees of freedom 
and the vector of mixing proportions for each component, respectively. 

\begin{table}[htbp]
	\centering
		\begin{tabular}{cccc}
			\hline 
				parameter	&	\proglang{R} arguments	&	Dimensions	&	Description \\
			\hline
				$\bmu$	&	\code{mu}	&	$p \times 1 \times g $	&	
						the location parameter \\	
				$\bSigma$	&	\code{sigma}	&	$p\times p\times g$	&	
						the scale matrix\\
				$\bdelta$	&	\code{delta}	&	$p\times 1\times g$	&	
						the skewness parameter\\	
				$\nu$	&	\code{dof}	&	$g\times 1$	&	the degrees of freedom \\						
				$\pi$	&	\code{pro}	&	$g\times 1$	&	
						the mixing proportions\\
			\hline
		\end{tabular}
	\caption{Structure of the model parameters in \pkg{EMMIX-uskew}.}
	\label{tab1}
\end{table}

The probability density function of a multivariate skew $t$-distribution
is calculated by the \code{dmst} function. The parameter \code{dat} 
is an $n \times p$ matrix, containing the coordinates of the $n$ point(s)
at which the density is to be evaluated. The following command will
return a vector of $n$ density values.
\begin{CodeInput}
dmst(dat, mu, sigma, delta, dof)
\end{CodeInput}

For a FM-uMST density, the function \code{dfmmst} can be used. 
\begin{CodeInput}
dfmmst(dat, mu, sigma, delta, dof, pro)
\end{CodeInput}

\subsection{Generating samples from a FM-uMST distribution}
\label{sec3.1}

Consider generating a random sample of $n$ $p$-dimensional 
uMST observations, with location parameter $\bmu$, 
scale matrix $\bSigma$, skewness parameter $\bdelta$, 
and degrees of freedom $\nu$. 
The function \code{rfmmst} supports two types of inputs 
-- the parameters can be passed as separate arguments, 
or as a single list argument \code{known} with elements 
as specified in Table~\ref{tab1}:       

\begin{CodeInput}
rfmmst(g, n, mu, sigma, delta, dof, pro, known=NULL, ...)
\end{CodeInput}

As an example, suppose that $\bmu = (1, 2)^T$, $\bSigma$ 
is the identity matrix, $\bdelta=(-1,1)^T$, and $\nu=4$. 
Then the following command will generate 
a random sample of $500$ observations from the 
$uMST_{2,4}(\bmu, \bSigma, \bdelta)$ distribution,
\begin{CodeInput}
R> rfmmst(1,500, c(1,2), diag(2), c(-1,1), 4, 1)
\end{CodeInput}

To generate a mixture of uMST random samples, the above command
can be issued. Alternatively, the parameters can be specified 
in a list structure (Table~\ref{tab1}) \code{obj} as follows: 
\begin{CodeInput}
R> obj <- list()
R> obj$mu <- list(c(17,19), c(5,22), c(6,10)) 
R> obj$sigma <- list(diag(2), matrix(c(2,0,0,1),2), matrix(c(3,7,7,24),2)) 
R> obj$delta <- list(c(3,1.5), c(5,10), c(2,0))
R> obj$dof <- c(1, 2, 3)
R> obj$pro <-  c(0.25, 0.25, 0.5)
R> rfmmst(3, 500, known=obj)
\end{CodeInput}

An output of the \code{rfmmst} function consists of $p+1$ columns.
The first $p$ columns are the coordinates of the generated sample. 
The last column indicates from which component 
each data point is generated.
Executing the above command will generate an output 
similar to the following:  
\begin{CodeOutput}
        [,1]       [,2]   [,3]
 [1,] 19.91520 20.48515      1
 [2,] 72.81161 33.41381      1
 [3,] 17.02193 23.29119      1
 [4,] 23.53926 19.27946      1
 [5,] 16.85195 21.21340      1
 [6,] 18.01906 18.16612      1
 [7,] 22.23609 21.12174      1
 [8,] 44.65444 28.23259      1
 [9,] 18.18883 26.72330      1
[10,] 20.18908 18.97005      1
     ... rest omitted ...
\end{CodeOutput}

\subsection{Fitting a single multivariate skew $t$-distribution}
\label{sec3.2}

To fit a specified FM-uMST model, the core function in \pkg{EMMIX-uskew}, 
\proglang{fmmst}, is used. This implements the algorithm described 
in Section~\ref{sec2}. A typical function call of \code{fmmst} is: 
\begin{Code}
fmmst(g, dat, initial=NULL, known=NULL, itmax=100, eps=1e-3, 
nkmeans=20, print=TRUE)
\end{Code}

The main arguments used within this function are:
\begin{itemize}
	\item \code{g}: a scalar that specifies 
			the number of uMST components to be fitted.
	\item	\code{dat}: an $n\times p$ matrix containing the data. 
	\item \code{initial}: a list that specifies the initial values 
			used to start the algorithm.
	\item \code{known}: a list that specifies any model parameters 
			that are known and so not required to be estimated. 
	\item \code{itmax}: a scalar that specifies 
			the maximum number of iterations to be used.
	\item \code{eps}:	a scalar that specifies the termination criterion of the 
			EM algorithm loop. 		
	\item \code{nkmeans}: an integer that specify the number of $k$-means trials 
			to be used to select the best set of initial values.
\end{itemize}
Note that if the initial values of the model parameters are 
provided by the user, the argument \code{initial} is expected 
to be structured as described in Table~\ref{tab1}. 
Similarly, \code{known} is expected to have the same structure. 
When \code{initial=NULL}, \proglang{fmmst} will generate a set of 
initial values using the procedure described in Section~\ref{sec2.2}. 
Any parameters specified in \code{known}
are taken as known parameters and hence are not estimated by \code{fmmst}.
There is no need to specify the values of all the parameters in 
\code{initial} and \code{known} when only 
some of the parameters are known. Parameters that are not specified 
in the function call are estimated by \code{fmmst}.
By default, \code{fmmst} performs 20 $k$-means attempts when searching 
for the best initial value. The user can specify a different value using \code{nkmeans}.  
The termination criterion for the \pkg{EMMIX-uskew} algorithm is controlled by
the parameters \code{itmax} and \code{eps}. The EM loop terminates when
either one of the two criterion is satisfied, whichever occurs first: 
(a) the EM loop reaches \code{itmax} iterations 
(default is 100 iterations), or 
(b) the absolute difference between the current log likelihood value and
that the asymptotic log lileklihood value is smaller than \code{eps} 
(default is \code{1e-3}). 
The last argument of \code{fmmst} is \code{print}.  
When the option \code{print} is set to \code{TRUE} (default), \proglang{fmmst} 
prints the log likelihood value at each iteration and displays a summary of
the parameters of the fitted model after termination. 
To turn off the print mode, simply set \code{print=FALSE}.
For further details of the arguments of \code{fmmst}, the reader 
is referred to the documentation of \code{fmmst}. 
This can be accessed by typing \code{?fmmst} at the \proglang{R} command prompt. 
 
We consider now the T-cell phosphorylation dataset \citep{J058} 
as an example of asymmetrically distributed data, available from \citet{S005}. 
The data contain measurements of blood samples stained 
with four antibodies, CD4, CD45RA, SLP76, and ZAP70. 
For illustration, we randomly select 500 observations
and focus on two of the variables, CD4 and ZAP70. 
To fit a MST model to this bivariate Lymphoma dataset,
under the default settings, the following command is issued:
\begin{CodeInput}
R> set.seed(12345)
R > data("Lympho")
R > LymphoSample <- Lympho[sample(1:nrow(Lympho), 500),]
R > Fit <- fmmst(1, LymphoSample)     
\end{CodeInput}

A summary of the output of the fitted model can be obtained 
using the \proglang{summary} function. This prints the values
of the fitted model parameters for each component. 
For a fitted uMST model, the weighting proportion (which is $1$)
is not printed.  
The following output shows a typical summary of a fitted 
single component uMST model. 
\begin{CodeOutput}
R > summary(Fit)
Finite Mixture of Multivariate Skew t-Distribution
with 1 component

Mean:
         [,1]
[1,] 4.808245
[2,] 5.500559


Scale matrix:
[[1]]
           [,1]       [,2]
[1,] 0.06778378 0.03721489
[2,] 0.03721489 0.04811898


Skewness parameter:
           [,1]
[1,] -0.7082174
[2,] -0.7990700


Degrees of freedom:
5.851434 


> summary(Fit)
Finite Mixture of Multivarate Skew t-Distributions
with  1  component

Mean:
         [,1]
[1,] 4.808245
[2,] 5.500559

Scale matrix:
[[1]]
           [,1]       [,2]
[1,] 0.06778378 0.03721489
[2,] 0.03721489 0.04811898

Skewness parameter:
           [,1]
[1,] -0.7082174
[2,] -0.7990700

Degrees of freedom:
5.851434 
\end{CodeOutput}

To view a more detailed output of the \proglang{fmmst} function,
the \code{print} function is called. This outputs a list
containing 11 elements. The first five elements give the estimates 
of the parameters of the fitted FM-uMST model, as described in 
Table~\ref{tab1}. 

The posterior probability of component membership is given by
the output argument \code{tau}, a $g\times n$ matrix
where the rows corresponds to the component number. 
The final partition of each data point, based on \code{tau},
is stored as \code{clusters}.
The value of the log likelihood function, evaluated with 
the current parameter estimates, is given by \code{loglik}. 
The last two arguments \code{aic} and \code{bic} are the values of
the Akaike information criterion (AIC) and 
the Bayes information criterion (BIC), respectively.
The following output shows an excerpt from the second part 
of the \code{print} output of the fitted model.

\begin{CodeOutput}
R> print(Fit)
Finite Mixture of Multivariate Skew t-Distributions
with 1 component
... first five components omitted ...

$tau
     [,1] [,2] [,3] [,4] [,5] [,6] [,7] [,8] [,9] [,10] [,11] [,12] [,13] [,14] 
[1,]    1    1    1    1    1    1    1    1    1     1     1     1     1     1    
... rest omitted ...

$clusters 
[1]    1    1    1    1    1    1    1    1    1     1     1     1     1     1   
... rest omitted ...

$loglik
[1] -880.7115

$aic
[1] 1777.423

$bic
[1] 1811.14
\end{CodeOutput}

As mentioned previously, initial values for the EM algorithm 
can be specified by the user.
Suppose an initial guess of $\bmu$ for the above example is $(5, 6)^T$, 
then one can specify $\bmu^{(0)}$ to be $(5, 6)^T$ by issuing the command:
\begin{CodeInput}
R> obj <- list()
R> obj$mu <- list(c(5, 6))
R> fmmst(1, LymphoSample, initial=obj)
\end{CodeInput}
This will start the EM algorithm with the specified value for $\bmu^{(0)}$, 
and the other parameters using (\ref{eq:init}). 
The user can further demand more $k$-means trials to be performed 
by increasing \code{nkmeans}, for example, to $50$ trials.
This can be achieved by issuing the following command.
\begin{CodeInput}
R> fmmst(1, LymphoSample, nkmeans=50)
\end{CodeInput}

\subsection{Fitting mixtures of multivariate skew $t$-distributions}
\label{sec3.3}

This section presents an illustration of fitting a mixture of 
unrestricted skew $t$-distributions to 
some bivariate bimodal asymmetric data. 
We consider the Australian Institute of Sport (AIS) data from
\citet{B007}, where thirteen body measurements on 102 male 
and 100 female athletes were recorded. 
In this example, we consider the clustering of the data
with a two component skew $t$-mixture model based on the two variables
Height and Body fat. By setting \code{print=TRUE}, we can examine 
the value of the log likelihood function at each iteration.    
 
\begin{CodeInput}
R> Fit2 <- fmmst(2, ais[,c(2,12)], print=TRUE) 
\end{CodeInput}

\begin{CodeOutput}
Finite Mixture of Multivariate Skew t-Distributions
with 2 components
  
  --------------------------------------------------
  Iteration  1 : loglik =  -1372.711 
  Iteration  2 : loglik =  -1370.495 
  Iteration  3 : loglik =  -1368.773 
  Iteration  4 : loglik =  -1367.392 
  Iteration  5 : loglik =  -1366.251 
       ... rest omitted ...
  --------------------------------------------------
  Iteration  100 : loglik =  -1343.541
     
  Component means:
            [,1]       [,2]
  [1,] 181.91720 181.448906
  [2,]  13.67975   5.814277

  Component scale matrices:
  [[1]]
           [,1]     [,2]
  [1,] 26.32964 18.75208
  [2,] 18.75208 16.12990

  [[2]]
            [,1]      [,2]
  [1,] 61.598537 2.3177048
  [2,]  2.317705 0.1515253

  Component skewness parameters:
            [,1]     [,2]
  [1,] -9.582015 3.591328
  [2,]  5.975328 5.729339

  Component degrees of freedom:
  60.03386 28.94895 

  Component mixing proportions:
  0.4102178 0.5897822      
\end{CodeOutput}

We compare the results with two other model-based clustering methods 
provided by the package \pkg{mixsmsn} \citep{S001} 
and \pkg{EMMIX-skew} \citep{S002}. 
As mentioned previously, this two models are based on mixture
of restricted versions of the multivariate skew $t$-distributions. 
The first model adopts the skew normal/independent 
skew $t$-distribution \citep{J066}
as its component densities,
which is equivalent to the restricted skew $t$-distribution \citep{J004}
used in the second model.  
However, it should be noted that, 
in the ECME algorithm implemented in the package \pkg{mixsmsn},
the component degrees of freedom are constrained to be the same.
A comparison of the table of cluster labels 
(permuted where necessary to minimize the number of misallocations)
with the true class labels
(given by \code{ais$Sex} in this example; M for male, F for female) 
reveals that the FM-uMST model has a higher number of correct allocations 
($183$ compared to $162$ and $157$ given by 
\pkg{mixsmsn} and \pkg{EMMIX-skew}, respectively). 
Thus, the unrestricted FM-uMST model in \pkg{EMMIX-uskew} 
gives a more accurate clustering in this case.
  
\begin{CodeInput}
R> library("mixsmsn")
R> Fit3 <- smsn.mmix(ais[c(2,12)], g=2, family="Skew.t", group=TRUE)
R> Fit4 <- EmSkew(ais[c(2,12)], 2, "mst", print=FALSE)
R> table(ais$Sex, Fit3$group)
     1  2
  M 91 11
  F 29 71
R> table(ais$Sex, Fit4$clust)
     1  2
  M 89 13
  F 32 68
R> table(ais$Sex, Fit2$clusters)  
     1  2
  M 97  5 
  F 14 86
\end{CodeInput}

\subsection{Testing for the significance of the skewness parameter}
\label{sec3.4b}

When we set $\bdelta = \bzero$ in (\ref{eq:MST}), 
we obtain the multivariate $t$-density.
The function 
\begin{CodeInput}
fmmt(g, dat, initial=NULL, known=NULL, itmax=100, eps=1e-3, nkmeans=20, 
print=TRUE)
\end{CodeInput} 
implements the EM algorithm for fitting finite mixtures of 
multivariate $t$ (FM-MT) distributions \citep{B003}. 

To test whether the skewness parameter in the FM-uMST model is significant, 
one can construct a likelihood ratio test for the null hypothesis 
$H_0: \bdelta_1 = \ldots = \bdelta_g = \bzero$ versus 
the alternative hypothesis where at least one of 
$\bdelta_h$ $(h=1, \ldots, g)$ is different from $\bzero$. 
This leads to the test statistic 
\begin{equation}
LR = - 2 \left(L_{t} - L_{st}\right),
\end{equation} 
where $L_t$ and $L_{st}$ denote the log likelihood value 
associated with the FM-MT model and the FM-uMST model, respectively. 
It follows that the test statistics is asymptotically distributed as 
$\chi^2_r$, where $r$ is the difference between the number of parameters 
under the alternative and null hypotheses. 
This test is implemented in the function
\code{delta.test(stmodel=NULL, tmodel=NULL, stloglik, tloglik, r)}, 
where the first two arguments are the output from 
\code{fmmst} and \code{fmmt} respectively. 
Alternatively, the user can provide the log likelihood values 
of the two models and the value of $r$ directly 
by specifying the last three arguments of \code{delta.test()}. 
The output of the function is the $P$-value of the test.   

Consider again the AIS example in Section \ref{sec3.4}. 
If we examine the cluster labels given by the FM-MT model, 
we can see that it yields a noticeably higher number of 
misallocations than the skew $t$-mixture model. 
A test for $\bdelta=\bzero$ can be performed by 
issuing the following commands.
In this case, the small $P$-value suggests there is strong evidence 
that the skewness parameter in the FM-uMST fit is 
significantly different from zero.   
\begin{CodeInput}
R> Fit5 <- fmmt(2, ais[,c(2,12)])
R> table(ais$Sex, Fit5$clusters)
     1  2
  M 77 23
  F 14 88
R> delta.test(Fit2, Fit5)
  0.0003798128
\end{CodeInput}

\subsection{Discriminant analysis}
\label{sec3.5}

Discriminant analysis based on a specified FM-uMST model 
can be performed using the \code{fmmstDA} function.
\begin{CodeInput}
fmmstDA(g, dat, model)
\end{CodeInput} 
The data in \code{dat} are assigned to the cluster 
corresponding to the component of the FM-uMST model 
with the highest posterior probability. 
Specifications of the model parameters must be provided
in \code{model}, which is typically an output from \code{fmmst}. 
Optionally, \code{model} can be specified by the user as 
a list of  at least six elements: the five model parameters, 
and a vector of cluster labels \code{clusters}.
The following commands shows an example using \code{fmmstDA}.
A random sample of FM-uMST variables is generated from \code{rfmmst},
the first part of which is used as training set, 
and the second is a testing set. 
The FM-uMST model fitted to the training set 
is then used for classifying the data in the testing set.       

\begin{CodeInput}
R> set.seed(732)
R> X <- rfmmst(3, 200, known=obj)
R> Ind <- sample(1:nrow(X),175)
R> train <- X[Ind,]
R> test <- X[-Ind,]
R> trainmodel <- fmmst(3, train[,1:2])
R> fmmstDA(3, test[,1:2], trainmodel)
R> results <- fmmstDA(3, test[,1:2], trainmodel)                            
R> table(test[,3], results)
   results
     1  2  3
  1  0  6  0
  2  0  0  5
  3 13  0  1   
\end{CodeInput}

\subsection{Visualization of fitted contours}
\label{sec3.4}

The \pkg{EMMIX-uskew} package supports visualization of the contours 
of a FM-uMST model in 2D and 3D. The plots are generated by the 
functions \code{fmmst.contour.2d} and \code{fmmst.contour.3d},
 
\begin{CodeInput}
fmmst.contour.2d(dat, model, grid=50, drawpoints=TRUE, clusters=NULL, 
	levels=10, component=NULL, map=c("scatter", "heat", "cluster"), ...)
fmmst.contour.3d(dat, model, grid=20, drawpoints=TRUE, clusters=NULL,
	levels=0.9, component=NULL, ...)
\end{CodeInput}

In \code{fmmst.contour.2d} (\code{fmmst.contour.3d}), 
the first argument \code{dat} is a matrix of coordinates 
with two (three) columns. The second argument \code{model},
similar to that in \code{fmmstDA()},
is either an output from \code{fmmst()},
or a list containing the five model parameters and the cluster labels.
The grid size is determined by \code{grid}. 
By default, the data points are included in the plot.
If only the contour are required, 
the option \code{drawpoints=FALSE} should be set. 
When including the points in a plot, \code{clusters} 
specifies the component labels of each point 
according to which the data points will be coloured.
The argument \code{levels} is either an integer specifying 
the number of contour lines to be plotted, or a vector of quantile values. 
For \code{fmmst.contour.3d}, only the $90$th
percentile contour is plotted by default. If more contours are required,
the argument \code{levels} should be a vector of the required quantiles.
For example, if a plot of the $25^{th}$, $50^{th}$, and $75^{th}$ percentiles 
are required, then \code{levels = c(0.25, 0.5, 0.75)}. 
Bivariate data have the option of being plotted as an intensity map 
instead of scatter plot. This can be obtained by setting \code{map="heat"}.
There is also an option for plotting a cluster map of a fitted model 
using the option \code{map="cluster"}.
Plots for specific components of a mixture model can be requested 
with the argument \code{component}. 
When \code{component=NULL} (which is default), the mixture contour 
is plotted. When \code{component} is a vector with length 
between $1$ and $g$, the specified components are plotted 
and the mixing proportion is not taken into account. 
The last argument of the \code{fmmst.contour} functions ``\code{...}" 
allows the user to pass additional arguments 
to the \code{plot} function, such as 
the colour and size of the points. \newline

\begin{figure}
	\centering
		\includegraphics[width=1.00\textwidth]{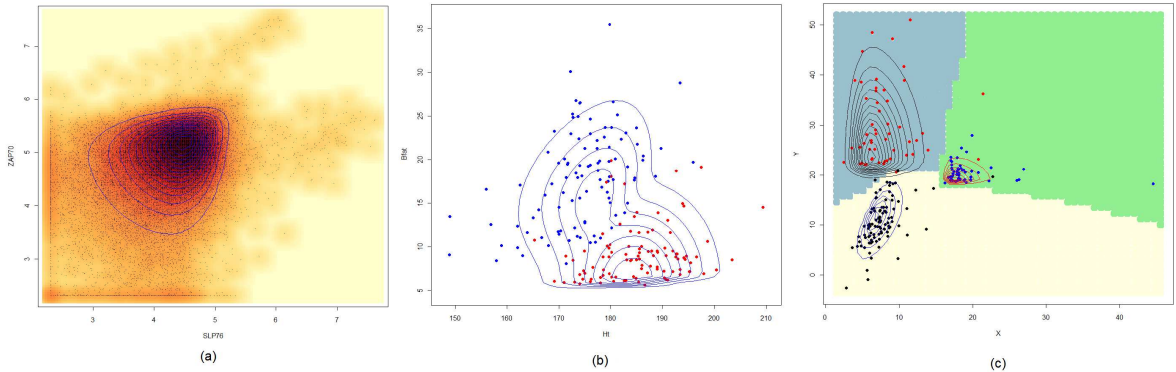}
	\caption{2D contour plots generated by the \code{fmmst.contour.2d}
		function. 
		(a) The fitted contour of the single component uMST model 
		plotted over the hue intensity diagram of the Lymphoma dataset;
		(b) the default mixture contour plot of the fitted 
		two-component FM-uMST model of the AIS dataset;
		(c) the contour of the individual components of the three-component 
		model fitted to a bivariate synthetic sample 
		plotted over the cluster map of the sample.}
	\label{fig1}
\end{figure}

Figure~\ref{fig1}a shows the contour of the fitted MST model
to the Lymphoma data. Here a heatmap of the original data is used.
This plot can be generated via the command,
\begin{CodeInput}
R> fmmst.contour.2d(Lympho, model=Fit, map="heat", 
   xlab="SLP76", ylab="ZAP70")
\end{CodeInput}

The default \code{fmmst.contour.2d} function will return 
a scatter plot of the data in 2D superimposed 
with the contours of the fitted mixture model. 
For example, the following command generates a contour plot of 
the fitted FM-uMST model to the \code{ais} data in Section~\ref{sec3.2}  
(Figure~\ref{fig1}b). 
Note that \code{fmmst.contour.2d} coloured the sample points 
according to the clustering given by the argument \code{clusters}: 
\begin{CodeInput}
R> label <- ais$Sex
R> label[label==0] <- 2
R> fmmst.contour.2d(ais[,c(2,12)], model=Fit2, clusters=label, 
	 xlab="Ht", ylab="Bfat")
\end{CodeInput}

Suppose we are interested in visualizing a clustering map   
of the fitted model to the simulated data in Section~\ref{sec3.5}.
This plot can be generated by issuing the following command. 
\begin{CodeInput}
R> fmmst.contour.2d(X, model=trainmodel, clusters=X[,3], map="cluster", 
		component=1:3)
\end{CodeInput}

The output is given in Figure~\ref{fig1}c.
\newline

To demonstrate the use of \code{fmmst.contour.3d}, 
we consider the clustering of a trivariate 
Diffuse Large B-cell Lymphoma (DLBCL) dataset 
provided by the British Columbia Cancer Agency.
The data contain fluorescent intensities of multiple
conjugated antibodies (known as markers) stained on
a sample of over 8000 cells derived from 
the lymph nodes of patients diagnosed with DLBCL.
In flow cytometric analysis, these parallel measurements
of fluorescent intensities can be used to study 
the differential expression of different surface 
and intracellular proteins of a given blood sample.   
The analysis typically involves the identification of 
cell populations from the multidimensional dataset,
currently performed manually by visually 
separating regions (gates) of interests
on a series of sequential bivariate projections of the data, 
a process known as \emph{gating}. 
Due to the subjective and time-consuming nature of this approach, 
and the difficulty in detecting higher-dimensional 
inter-marker relationships,   
many efforts have been made to develop 
computational methods to automate the gating process.

The DLBCL samples here were stained with three markers CD3, CD5, and CD19. 
The task is to automatically gate the cells
by clustering the data into four groups.
Hence we fit a four-component FM-uMST model to the data.
The maximum number of iterations was increased to $300$.

\begin{figure}
	\centering
		\includegraphics[width=0.75\textwidth]{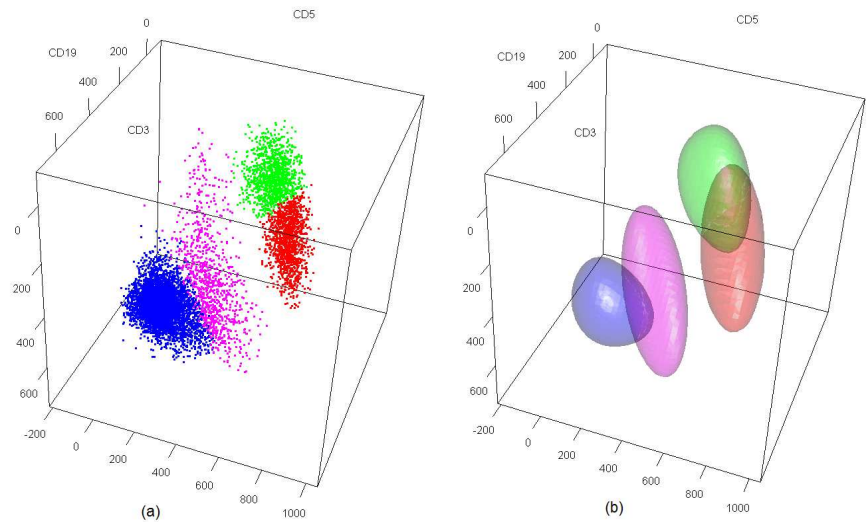}
	\caption{3D contours plot of the DLBCL dataset 
		generated by the \code{fmmst.contour.3d} function. 
		(a) A scatterplot of the data coloured according 
		to the the true clustering labels of the DLBCL dataset;
		(b) fitted contour of the three component FM-uMST model 
		for the DLBCL dataset.}
	\label{fig2}
\end{figure}

A scatterplot of the data is shown in Figure~\ref{fig2},
where the dots are coloured according to the clustering
provided by human experts, which are considered as the `true' class labels. 
Figure~\ref{fig2}b shows the $95^{th}$ percentile density 
contours of the four components of the fitted model
which are displayed with matching colours.
The 3D plot uses the \pkg{rgl} visualization device system, 
and hence supports user friendly interactive navigation.
The plots can be rotated in real-time to select a suitable 
viewpoint. The following code can be used to generate the 
3D plots in Figure~\ref{fig2}. 
 
\begin{CodeInput}
R> Fit6 <- fmmst(DLBCL, 4, itmax=300)
R> fmmst.contour.3d(DLBCL, model=Fit4, level=0.9, drawpoints=FALSE, 
   xlab="FL1.LOG", ylab="FL2.LOG", zlab="FL4.LOG", quantile=0.95)
\end{CodeInput}
 
The effectiveness of a clustering can be obtained 
by comparing its error rate with the cluster labels 
from manual expert gating taken to be the true class labels. 
This error rate is calculated for each permutation 
of the cluster labels of the clustering result under consideration 
and the rate reported is the minimum value over all such permutations.
Note that dead cells were removed before evaluating 
the error rate against the benchmark results. 
For comparison, we calculated the error rate associated with
the clustering results given by three other methods -- 
FLAME \citep{J004}, flowClust \citep{J069} and flowMeans \citep{J070}. 
From Table~\ref{tab2}, the FM-uMST model clearly shows superior performances
in this dataset. \newline

\begin{table}
	\centering
		\begin{tabular}{|c|c|c|c|c|}
				\hline
				Method	&	FLAME	&	flowClust	&	flowMeans	&	FM-uMST	\\
				\hline
				Error rate	&	0.074	&	0.076	&	0.140	&	0.046	\\
				\hline	
		\end{tabular}
	\caption{Error rate of misclassification of four methods for the DLBCL dataset.}
	\label{tab2}
\end{table}

\section{Concluding remarks}
\label{sec4}

We have presented the \proglang{R} package \pkg{EMMIX-uskew} 
for fitting finite mixtures of unrestricted multivariate
skew $t$-distributions to heterogeneous asymmetric data.
The package implements a closed-form EM algorithm for fitting 
FM-uMST models and provides user-friendly visualization 
of the fitted contours in 2D and 3D. The major features 
of the software have been demonstrated on three real examples 
on the T-cell phosphorylation data, 
the Australian Institute of Sports (AIS) data, and the DLBCL dataset. 
The clustering results were compared to those obtained via 
mixtures of restricted multivariate skew $t$-distributions 
and other methods.
In both the AIS and DLBCL illustrations, the unrestricted model
gave better clustering results with respect to the true class labels.

It should be noted that the fitting of 
the unrestricted skew $t$-mixture model
can be quite slow in higher dimensional applications, 
due to the computationally intensive procedure
involved in the calculation of multivariate $t$-distribution function values.
The algorithm would benefit from further research on 
applicable acceleration techniques, for example, 
the implementation of the SQUAREM strategy \citep{J124}.

\section*{Acknowledgments}
This work is supported by a grant from the Australian Research Council.
Also, we would like to thank Professor Seung-Gu Kim for comments and corrections, 
and Drs. Kui (Sam) Wang, Saumyadipta Pyne, and Felix Lamp 
for their helpful discussions on this topic.

\bibliography{Refs}

\end{document}